\documentclass[aps,prb,a4paper,twocolumn,floatfix,showpacs,preprintnumbers]{revtex4-1}

\usepackage{amsmath}
\usepackage{amssymb}
\usepackage{graphicx}
\usepackage{color}

\begin{document}

\title{Influence of correlated impurities on conductivity of graphene
sheets: Time-dependent real-space Kubo approach}
\author{T. M. Radchenko}
\affiliation{Solid State Electronics, Department of Science and Technology (ITN), Link\"{o}ping
University, 60174 Norrk\"{o}ping, Sweden}
\affiliation{Deptartment of Solid State Theory, Institute for Metal Physics, NASU,
36 Acad. Vernadsky Blvd., 03680 Kyiv, Ukraine}
\author{A. A. Shylau}
\affiliation{Solid State Electronics, Department of Science and Technology (ITN), Link\"{o}ping
University, 60174 Norrk\"{o}ping, Sweden}
\author{I. V. Zozoulenko}
\affiliation{Solid State Electronics, Department of Science and Technology (ITN), Link\"{o}ping
University, 60174 Norrk\"{o}ping, Sweden}
\email{igor.zozoulenko@itn.liu.se}
\date{\today }

\begin{abstract}
Exact numerical calculations of the conductivity of graphene sheets with
random and correlated distributions of disorders have been performed using
the time-dependent real-space Kubo formalism. The disorder was modeled by
the long-range Gaussian potential describing screened charged impurities and
by the short-range potential describing neutral adatoms both in the weak and
strong scattering regime. Our central result is that correlation in the
spatial distribution for the strong short-range scatterers and for the
long-range Gaussian potential do not lead to any enhancement of the
conductivity in comparison to the uncorrelated case. Our results strongly
indicate that the temperature enhancement of the conductivity reported in
the recent study (Yan and Fuhrer, Phys. Rev. Lett. 107, 206601 (2011)) and
attributed to the effect of dopant correlations was most likely caused by
other factors not related to the correlations in the scattering potential.
\end{abstract}

\maketitle

\section{Introduction}

Investigation of charge transport in graphene and understanding factors that
affect its conductivity represent one of the central directions of graphene
research.\cite{CastroNetoreview,PeresReview,DasSarmaReview} This is
motivated by both fundamental interest to graphene's transport properties as
well as by potential applications of this novel material for electronics. It
is commonly recognized that the major factors determining the electron
mobility in graphene are long-range charged impurities trapped on the
substrate and strong resonant short-range scatterers due to adatoms
covalently bound to graphene.\cite{PeresReview} The nature of scatterers is
directly reflected in the dependence of the conductivity on the electron
density, $\sigma =\sigma (n)$, and therefore investigation of this function
constitutes the focus of experimental and theoretical research.\cite%
{PeresReview,DasSarmaReview} For example, both the strong short-range
potential\cite{Stauber,Katsnelson,Ostrovsky,Klos,Yuan10,Ferreira} and the
long-range potential \cite{Ando2006,Nomura2006,Hwang2007,Klos} lead to
similar linear density dependence of the conductivity commonly observed in
experiments.\cite%
{Novoselov2005,Tan2007,Morozov2008,Ponomarenko2009,Monteverde2010,Ni2010}.
Experiments also often show strong deviations from this linear dependence
and an asymmetry with respect to the polarity of carriers, which can be
attributed to a number of factors such as the effect of realistic structural
defects and impurities\cite%
{Robinson,Wehling,Yuan10,Leconte11,Lherbier12,Mayou}, effect of contacts\cite%
{GoldhaberGordon,Lopez,Huang}, effect of weak short-range impurities\cite%
{Nomura2006,Ando2006,Hwang2007,Stauber}, the ballistic or quasiballistic
transport regimes\cite{Du,BolotinPRL,Klos2009,Klos} and many others.

Detailed studies of the density dependence of the conductivity in graphene
often require exact numerical approaches for transport calculations combined
with \textit{ab initio} calculations for microscopic properties of realistic
scatterers.\cite{Robinson,Wehling,Yuan10,Leconte11,Lherbier12} Such the
approaches can capture the essential features of the system at hand as well
as can address transport regimes which are not accessible by other means.
Exact numerical approaches can also be used to test a validity of
conclusions of various semi-classical analytical approaches and
applicability of approximations used in such the approaches. Among the most
popular numerical methods reported in the literature are the recursive
Green's function technique\cite{Lewenkopf,Xu,Ihnatsenka} and the
time-dependent real space Kubo method\cite%
{Roche1997,Triozon2002,Triozon2004,Markussen,Ishii,Yuan10,Ferreira,Leconte11,Lherbier12,Mayou}%
. The later method is especially suited to treat large graphene systems
containing tens of millions of atoms with dimensions approaching realistic
systems.

Recently, Yan and Fuhrer \textit{}\cite{YanFuhrer} addressed the
effect of correlation in the spatial distribution of disorder on the
conductivity of graphene sheet by doping it with potassium atoms. They found
that the conductivity of the system at hand increases as the temperature
rises, and argued that this was caused by the enhancement of correlation
between the potassium ions due to the Coulomb repulsion. This conclusion, in
turn, was based on the theoretical predictions of Li \textit{et al.}\cite{Li}
that the correlations in the position of long-range scatterers strongly
enhances the conductivity. It should be noted that the approach of Li
\textit{et al.}\cite{Li} is based on the standard Boltzmann approach within
the Born approximation. However, the applicability of the Born approximation
for graphene has been questioned in Ref. \cite{Klos} where it was shown that
predictions based on the standard semiclassical Boltzmann approach within
the Born approximation for the case of the long-range Gaussian potential are
in quantitative and qualitative disagreement with the exact
quantum-mechanical results in the parameter range corresponding to realistic
systems. (A discussion of various aspects of the applicability of the Born
approximation for graphene and bilayer graphene can be found in Refs. \cite%
{Ferreira,Klos,Xu2011}). It is therefore not clear to what extend the
semiclassical predictions of Li \textit{et al.}\cite{Li} based on the Born
approximation are justified for the case of correlated impurities. Since the
conclusions of the experiment of Yan and Fuhrer \textit{}\cite%
{YanFuhrer} rely essentially on the above semi-classical predictions, it is
of interest to study the effect of spatial correlation of disorders by exact
numerical methods.

The main aim of the present paper is therefore to investigate the influence
of the spatial correlation of disorder on the conductance of graphene sheets
using an exact quantum mechanical approach. In the present study we utilize
the time-dependent real-space quantum Kubo method\cite%
{Roche1997,Triozon2002,Triozon2004,Markussen,Ishii,Yuan10,Ferreira,Leconte11,Lherbier12,Mayou}%
. allowing us to study large graphene sheets containing millions of atoms.
We consider models of disorder appropriate for realistic impurities,
including the Gaussian potential describing screened charged impurities and
the short-range potential describing neutral adatoms. The paper is organized
as follows. The basic model of the system under consideration including
models for impurity potentials as well as the basics of the numerical Kubo
method are formulated in Section II. Section III presents and discusses the
obtained results and compares our numerical findings with available
experimental data and theoretical predictions. The conclusions are
summarized in Section IV. A various details of the numerical approach are
presented in the Appendix.

\section{Tight-binding model and time-dependent real-space Kubo formalism}

\begin{figure}[tbh]
\includegraphics[keepaspectratio, width=0.6\columnwidth]{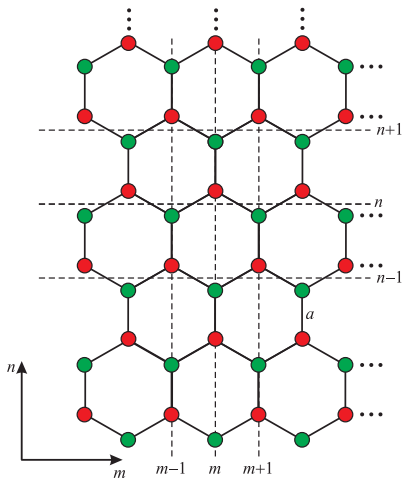}
\caption{(Color online) A honeycomb graphene lattice. }
\label{fig:graphene}
\end{figure}

We model electron dynamics in graphene using the standard $p$-orbital
nearest neighbor tight-binding Hamiltonian defined on a honeycomb lattice%
\cite{PeresReview,DasSarmaReview}, Fig. \ref{fig:graphene}.%
\begin{equation}
\hat{H}=-u\sum_{i,i^{\prime }}c_{i}^{\dagger }c_{i^{\prime
}}+\sum_{i}V_{i}c_{i}^{\dagger }c_{i},  \label{H}
\end{equation}%
where $c_{i}^{\dagger }$ and $c_{i}$ are the standard creation and
annihilation operators acting on a quasiparticle on the site $i=(m,n)$. The
summation over $i$ runs over the entire graphene lattice, while $i^{\prime }$
is restricted to the sites next to $i$; $u=2.7$ eV is the hopping integral
for the neighboring atoms $i$ and $i^{\prime }$, and $V_{i}$ is the on-site
potential describing impurity scattering.

In the present study we consider both short- and long-range impurities. The
short-range impurities represent neutral adatoms on graphene and are modeled
by the delta-function potential%
\begin{equation}
V_{i}=\sum_{j=1}^{N_{imp}}V_{j}\delta _{ij},  \label{delta}
\end{equation}%
where $N_{imp}$ is the number of impurities on a graphene sheet. Estimations
based on \textit{ab initio} calculations and the T-matrix approach for
common adatoms provide typical values for the on-site potential $%
V_{j}=V_{0}\lesssim 80u.$\cite{T-matrix,Robinson,Wehling,Ihnatsenka,Ferreira}
(For example, for hydrogen, $V_{0}\approx 60u)$.

The long-range impurities corresponds to charged ions situated typically on
a surface of a dielectric substrate. We model them by the Gaussian potential
commonly used in the literature,\cite%
{PeresReview,DasSarmaReview,Klos,Adam,Xu2011}%
\begin{equation}
V_{i}=\sum_{j=1}^{N_{imp}}U_{j}\exp \left( \frac{|\mathbf{r}_{i}-\mathbf{r}%
_{j}|}{2\xi ^{2}}\right) ,  \label{Gauss}
\end{equation}%
where $\mathbf{r}_{i}$ is the radius-vector of the site $i,$ $\xi $ is the
effective screening length, and the potential height is uniformly
distributed in the range $U_{j}\in \lbrack -\Delta ,\Delta ]$ with $\Delta $
being the maximum potential height.

\begin{figure}[tbh]
\includegraphics[keepaspectratio, width=1\columnwidth]{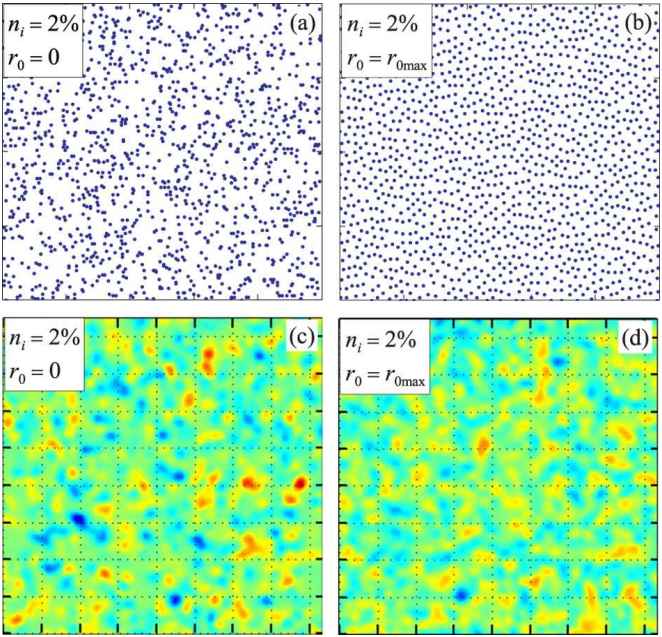}
\caption{(Color online) A representative illustration of random and
correlated distributions of impurities for the cases of (a), (b) short-range
impurities, and (c), (d) Gaussian impurities for the impurity concentration $%
n_{i}=2\%$.}
\label{fig:RandomImpurities}
\end{figure}

\qquad We consider two cases of impurity distribution, random (uncorrelated)
and correlated. In the first case of uncorrelated impurities the summation
in Eqs. (\ref{delta}), (\ref{Gauss}) is performed over the random
distribution of impurities over the honeycomb lattice. In the second case
impurities are no longer considered to be randomly distributed and to
describe their spatial correlation we adopt a model used in Ref. \cite{Li}
introducing the pair distribution function $g(r)$,
\begin{equation}
g(r)=\left\{
\begin{array}{c}
0,\;r<r_{0}\text{;} \\
1,\;r>r_{0}\text{,}%
\end{array}%
\right.   \label{g(r)}
\end{equation}%
where $r$ is the distance between two impurities and the correlation length $%
r_{0}$ defines the minimum distance that can separate two impurities. (Note
that for the randomly distributed (uncorrelated) impurities $r_{0}=0$). The
largest distance $r_{0_{\max }}$ depends on the relative impurity
concentration $n_{i};$ the smaller the concentration $n_{i},$ the larger $%
r_{0_{\max }}$. Calculated values of $r_{0_{\max }}$ for the different
relative impurity concentrations $n_{i}$ are presented in Table I, and
representative examples of random and correlated distributions for the cases
of the short- and long-range potentials are shown in Fig. 2.

\begin{table}[t]
\begin{tabular}{|l|l|l|l|l|l|l|}
\hline
$n_{i}$ & 0.5\% & 1\% & 2\% & 3\% & 4\% & 5\% \\ \hline
$r_{0_{\max }}(a)$ & 13 & 9 & 6 & 5 & 4 & 3 \\ \hline
\end{tabular}%
\caption{Relation between the relative concentration of impurities $n_{i}$
and the largest correlation distance $r_{0_{\max }}$ (expressed in units of
the carbon-carbon distance $a=0.142$ nm).}
\end{table}

The transport properties of graphene sheets are calculated on the basis of
the time-dependent real-space Kubo formalism where the dc conductivity $%
\sigma $ is extracted from the wave packet temporal dynamics governed by the
time-dependent Schr\"{o}dinger equation\cite%
{Roche1997,Triozon2002,Triozon2004,Markussen,Ishii}. This is a
computationally efficient method scaling with a number of atoms in the
system $N$, and thus allowing treating very large graphene sheets containing
many millions of C atoms.\cite{Yuan10,Leconte11,Lherbier12,Ishii,Mayou}

The calculation of the dc conductivity starts from the Kubo-Greenwood
formula, \cite{Madelung}%
\begin{equation}
\sigma =\frac{2\hbar e^{2}\pi }{\Omega }\text{Tr}\left[ \hat{v}_{x}\delta (E-%
\hat{H})\hat{v}_{x}\delta (E-\hat{H})\right] ,  \label{Kubo-Greenwood}
\end{equation}%
where $\hat{v}_{x}$ is the $x$-component of the velocity operator, $E$ is
the Fermi energy, $\Omega $ is the area of the graphene sheet, and factor 2
accounts for the spin degeneracy. Let us introduce the mean quadratic
spreading of a wave packet along the $x$-direction at the energy $E$, $%
\left\langle \Delta \hat{X}^{2}(E,t)\right\rangle =\left\langle \left( \hat{X%
}(t)-\hat{X}(0)\right) ^{2}\right\rangle $ where $\hat{X}(t)=\hat{U}%
^{\dagger }(t)\hat{X}\hat{U}(t)$ is the position operator in the Heisenberg
representation, and $\hat{U}(t)=e^{-i\hat{H}t/\hbar }$ is the time-evolution
operator. The conductivity can then be expressed as the Einstein relation%
\begin{equation}
\sigma \equiv \sigma _{xx}=e^{2}\tilde{\rho}(E)\lim_{t\rightarrow \infty
}D(E,t),  \label{sigma(t)}
\end{equation}%
where $\tilde{\rho}(E)=\rho /\Omega =$Tr$\left[ \delta (E-\hat{H})\right]
/\Omega $ is the density of sates (DOS) per unit area (per spin), and the
time-dependent diffusion coefficient $D(E,t)$ is related to $\Delta \hat{X}%
^{2}(E,t),$\cite{derivative}
\begin{subequations}
\label{Dx}
\begin{gather}
D(E,t)=\frac{\left\langle \Delta \hat{X}^{2}(E,t)\right\rangle }{t}
\label{a} \\
=\frac{1}{t}\frac{\text{Tr}\left[ \left( \hat{X}_{H}(t)-\hat{X}(0)\right)
^{2}\delta (E-\hat{H})\right] }{\text{Tr}\left[ \delta (E-\hat{H})\right] }.
\label{b}
\end{gather}%
Hence, calculation of $\sigma $ using the time-dependent real-space Kubo
formalism requires numerical evaluation of the mean quadratic spreading of
wave packets as prescribed by Eqs. (\ref{sigma(t)})$-$(\ref{Dx}). Details of
numerical calculations of $\sigma $ and numerical solution of the
time-dependent Schr\"{o}dinger equation are given in the Appendix.

\begin{figure}[tbh]
\includegraphics[keepaspectratio, width=1\columnwidth]{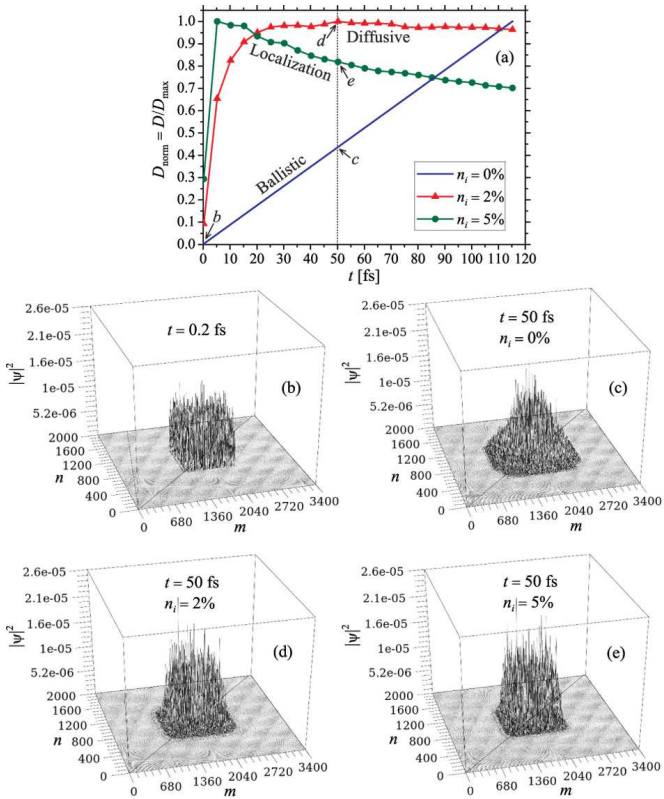} %
\caption{(Color online) (a) Temporal dependence of the diffusion coefficient
for different concentration of strong short-range impurities. The on-site
potential $V\sim 37u;$ $E=0.2u.$ (b)$-$(e) Wave packet propagation in
graphene lattice with different concentration of short-range impurities. (b)
Initial random state of the size 1020$\times $600 sites given by Eq. (%
\protect\ref{psiRandom}). (c)$-$(e) The wave packet shape after $t=50$ fs
for different impurity concentrations corresponding to ballistic, diffusive
and localization regimes from (a).}
\label{fig:diff}
\end{figure}

The diffusion coefficient $D(E,t)$ defined according to Eq. (\ref{Dx}) is
generally time-dependent and in different time intervals exhibits different
temporal behavior depending on whether the transport regime is (i)
ballistic, (ii) diffusive and (iii) localized.\cite{Leconte11,Lherbier12}
This is illustrated in Fig. \ref{fig:diff} showing a temporal evolution of the
diffusion coefficient for a graphene sheet with different impurity
concentrations $n_{i}$. The corresponding shapes of the wave packets for the
different transport regimes are visualized in Fig. \ref{fig:diff} for a some
representative time $t=50$ fs. In a system with no impurities ($n_{i}=0)$
electrons propagate ballistically such that the mean spreading of the wave
packet is $\sqrt{\left\langle \Delta \hat{X}^{2}(E,t)\right\rangle }\simeq
v_{F}t,$ where $v_{F}$ is the electron Fermi velocity. As a result, the
diffusion coefficient increases linearly with time, $D\simeq v_{F}^{2}t,$
with the slope being given by $v_{F}^{2}.$ In systems with a finite impurity
concentration the ballistic regime lasts up to times $\tau _{ball}\simeq
d_{i}/v_{F}$ where $d_{i}=n_{i}^{-1/2}$ is the average distance between
impurities.

For times $t>\tau _{ball}$ the system enters the classical diffusive regime
when the diffusion coefficient becomes independent of time. For the impurity
concentration $n_{i}=2\%$ in Fig. \ref{fig:diff} this regime starts at the time
$t\approx 25$ fs when the diffusion coefficient $D$ reaches its saturation.
For larger times the diffusion coefficient starts to decrease due to quantum
effects leading to the weak or the strong localization. This is in contrast
to classical diffusion where $D$ does not change when $t\rightarrow \infty .$
For example, for $n_{i}=5\%$ the localization effects becomes dominant
already at $t\gtrsim 10$ fs when the diffusion coefficient starts to
decrease with time, see Fig. \ref{fig:diff}. (Note that for $n_{i}=2\%$ the
decrease of $D$ takes place at times exceeding the time interval shown in
the figure). It should be stressed that in the present study we are
interested in the diffusive transport regime when the diffusion coefficient
reaches its maximum. Therefore, following Refs. \cite{Leconte11,Lherbier12},
we replace in Eq. (\ref{sigma(t)}) $\lim_{t\rightarrow \infty
}D(E,t)\rightarrow D_{\max }(E),$ such that the dc conductivity is defined
as
\end{subequations}
\begin{equation}
\sigma =e^{2}\tilde{\rho}(E)D_{\max }(E).  \label{sigmaMax}
\end{equation}%
It is noteworthy that within the Boltzmann approach the conductivity is
given by $\sigma _{Boltz}=e^{2}\tilde{\rho}(E)\tau \frac{v_{F}^{2}}{2}$,
where $\tau $ is the scattering time. Hence, it follows from Eq. (\ref%
{sigmaMax}) that the elastic length is related to the computed diffusion
coefficient via $l_{e}=v_{F}\tau =2D_{\max }/v_{F}.$

In most experiments the conductivity is measured as a function of the
electron density $n.$ Having calculated the DOS $\tilde{\rho}(E)$ as
described in the Appendix we compute the electron density $%
n(E)=\int_{-\infty }^{E}\tilde{\rho}(E)dE-n_{\text{ions}},$ where $n_{\text{%
ions}}=3.9\times 10^{15}$cm$^{-2}$ is the density of the positive ions in
the graphene lattice compensating the negative charge of the $p$-electrons
(note that for the ideal graphene lattice at the neutrality point $n(E)=0).$
(A dependence $n=n(E)$ is illustrated in Fig. \ref{fig:DOS} for different
scattering potentials). Combining the calculated $n(E)$ with the
conductivity $\sigma (E)$ given by Eq. (\ref{sigmaMax}), we compute the
density dependence of the conductivity, $\sigma =\sigma (n).$

\section{Results and discussion}

In this section we present results for the numerical conductivity of the
graphene sheets with random and correlated impurities described by the
short- and long-range potentials (Eqs. (\ref{delta}) and (\ref{Gauss})
respectively).

\subsection{Short-range impurities}

Depending on the impurity strength, the conductivity of graphene sheets is
expected to exhibit qualitatively different density dependence for the weak
and strong short-range scattering. The standard Boltzmann approach within
the Born approximation predicts that the conductivity is independent on the
electron density, \cite{PeresReview,DasSarmaReview,Stauber,Ferreira}%
\begin{equation}
\sigma =\frac{8e^{2}}{h}\frac{\left( v_{F}\hbar \right) ^{2}}{n_{i}V_{0}^{2}}%
,  \label{Boltzmann}
\end{equation}%
where $v_{F}$ is the Fermi velocity in graphene, $v_{F}\hbar =\frac{3}{2}ua,$
and $V_{0}$ is the strength of the on-site potential in Eq. (\ref{delta}), $%
V_{j}=V_{0}$. Going beyond the Born approximation one obtains that strong
scatterers, with the accuracy up to logarithmic corrections, lead to a
linear density dependency,\cite{Stauber,Katsnelson,Ostrovsky}%
\begin{equation}
\sigma =\frac{4e^{2}}{h}\frac{n}{n_{i}}\left( \ln \sqrt{\pi n}R_{0}\right)
^{2},  \label{Stauber}
\end{equation}%
where $R_{0}$ is the scatterer's radius. Note that the condition for the
validity of the Born approximation in Eq. (\ref{Boltzmann}), can be
presented in the form,\cite{Ferreira,Xu2011} $\left( |V_{0}|/u\right)
^{-1}\ll \sqrt{n_{e}}\ln n_{e},$ where $n_{e}$ is the relative electron
density (i.e. the electron density per C atom). For realistic electron
densities, $n_{e}\lesssim 0.01,$ this condition loosely corresponds to $%
|V_{0}|$ $\lesssim u,$ which we adopt as a definition of the weak impurity
scattering. The condition $|V_{0}|$ $\gg u$ (when Eq. (\ref{Stauber}) is
expected to be valid), defines the case of the strong impurity scattering.
Note that most of adatoms present in graphene falls into the second case of
the strong scattering with the effective on-site potentials in the range $%
|V_{0}|/u\approx 10-80$.\cite{T-matrix}

\subsubsection{Short-range potential, strong scattering, $|V_{0}|$ $\gg u$}

\begin{figure*}[tbh]
\includegraphics[keepaspectratio, width=2.0\columnwidth]{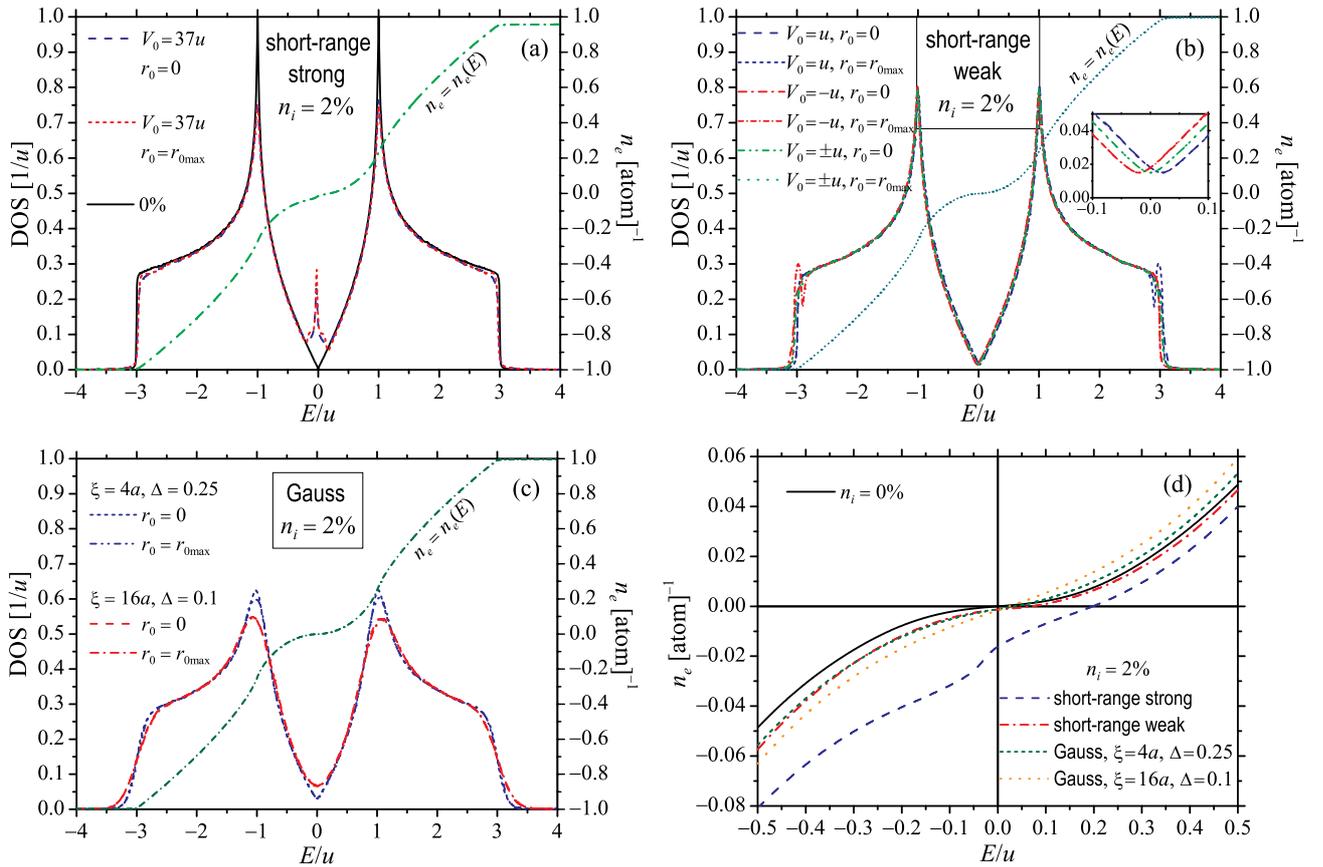}
\caption{(Color online) Density of states (DOS) and the relative charge
carrier concentration $n_e$ (the number of electrons per C-atom) as function
of the energy $E$ for (a) short-range strong, (b) short-range weak, and (c)
Gaussian potential for random and correlated impurities. (c) Dependencies $%
n=n(E)$ for short- and long-range potentials.}
\label{fig:DOS}
\end{figure*}

Figures \ref{fig:DOS} (a), (d) show the DOS and the electron density $n=n(E)$
in a graphene sheet with strong scatterers, $V_{0}=37u$ and the
concentration $n_{i}=2\%$. The calculated dependencies are practically
identical for the cases of random and correlated distributions of
impurities. The calculated DOS shows a pronounced peak in the vicinity of
the Dirac point, $-0.3\lesssim \frac{E}{u}\lesssim 0.1$ which reflects a
formation of the impurity band as discussed in Ref. \cite{Yuan10}. Note that
the peak is shifted with respect to $E=0$ which is related to the asymmetry
of the impurity potential ($V_{0}>0).$ The shift of the impurity band also
leads to the asymmetry in $n=n(E)$ dependence, see Fig. \ref{fig:DOS} (d).

\begin{figure*}[tbh]
\includegraphics[keepaspectratio, width=1.8\columnwidth]{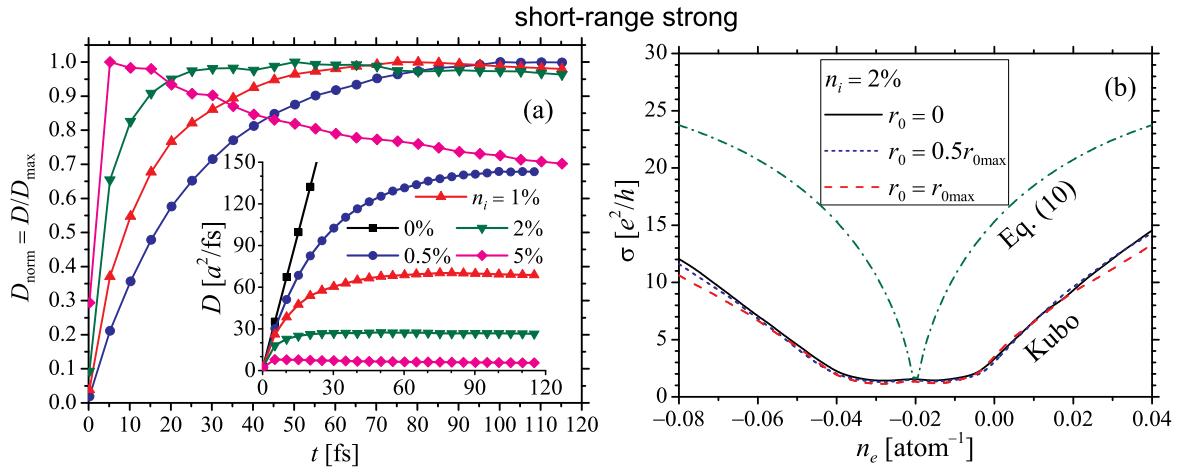}
\caption{(Color online) (a) Diffusion coefficient $D=D(t)$ for different
concentration of the short-range strong impurities; $V_{0}=37u$, $E=0.2u$.
(b) The conductivity $\protect\sigma $ as a function of the relative
electron density $n_{e}$ (the number of electrons per C-atom) for random and
correlated impurities, $n_{i}=2\%$. The dot-dashed line corresponds to Eq.
(10), which is shifted to the charge neutrality point at $n_{e}\approx 0.02$.
}
\label{fig:sigmaSRstrong}
\end{figure*}

Figure \ref{fig:sigmaSRstrong} (a) shows the time-dependent diffusion
coefficient for different impurity concentrations for the case of random
distribution of impurities. The temporal dependencies $D=D(t)$ demonstrate
that the system at hand reaches and stays in the diffusive transport regime
for the impurity concentration $n_{i}=1\%$ and $2\%$ at times $t\gtrsim 70$
and 30 fs correspondingly. For the impurity concentration $n_{i}=0.5\%$ the
diffusive regime is reached at times outside those shown in the figure ($%
t\gtrsim 130$ fs), whereas for $n_{i}=5\%$ the system almost immediately
reaches the localization regime.

To analyze the density dependence of the conductivity we choose a
representative concentration $n_{i}=2\%$ and $t=80$ fs when $D(t)$ exhibits
a saturation corresponding to a well-defined diffusive regime, see Fig. \ref%
{fig:sigmaSRstrong} (b). The randomly distributed impurities show a
quasi-linear density dependence of the conductivity. It is worth to note
that the calculated conductivity fully reproduces numerical results reported
by Yuan \textit{et al.}\cite{Yuan10} who used a similar time-dependent Kubo
approach. The corresponding theoretical prediction exhibits pronounced
deviations from the linear dependence caused by logarithmic corrections in
Eq. (\ref{Stauber}). These deviations are much stronger than those typically
seen experimentally.\cite{PeresReview} This is related to the fact that in
realistic experimental samples the electron densities are lower than those
used in our calculations,\cite{ElDensity} such that at smaller densities Eq.
(\ref{Stauber}) also exhibits a quasi-linear dependence. In the vicinity of
the Dirac point the conductivity flattens out which is related to a
transport regime due to a formation of the impurity band. Apparently, the
theoretical prediction Eq. (\ref{Stauber}) does not reproduce this transport
regime. For larger densities away from the Dirac point the calculated Kubo
conductivity differs by a factor of $\sim 2$ from the theoretical
predictions given by Eq. (\ref{Stauber}). Note that the calculated
dependence $\sigma =\sigma (n)$ is shifted with respect to $n=0,$which is
caused by the asymmetry in the dependence $n=n(E)$ due to the impurity band
as discussed above.

Figure \ref{fig:sigmaSRstrong} (b) also shows the dependence $\sigma =\sigma
(n)$ for the case of correlated impurities with the correlation lengths $%
r_{0}=0.5r_{0\max }$ and $r_{0}=r_{0\max }.$The central result is that
correlation in the impurity distribution practically does not affect the
conductivity of the system even for the largest correlation length $r_{0\max
}.$

\subsubsection{Short-range potential, weak scattering, $|V_{0}|$ $\lesssim u$%
.}

\begin{figure*}[tbh]
\includegraphics[keepaspectratio, width=1.8\columnwidth]{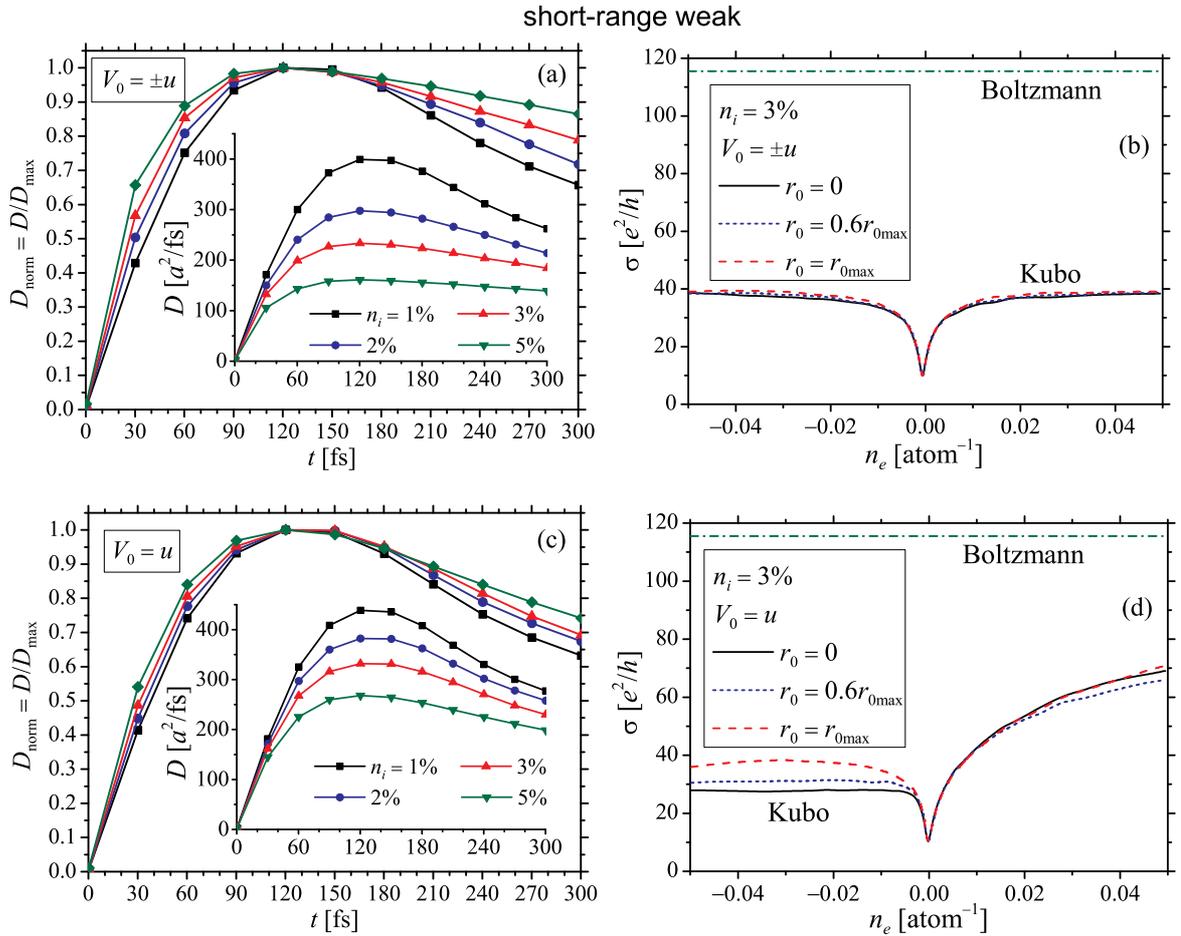}
\caption{(Color online) (a), (c) The diffusion coefficient $D=D(t)$ for for
short-range weak impurities with different concentrations $n_{i}$; $E=0.2u$.
(b), (d) The conductivity $\protect\sigma $ as functions of the relative
electron density $n_{e}$ (the number of electrons per C-atom) for random and
correlated short-range weak impurities, $n_{i}=3\%$. Dot-dashed lines shows
the Boltzmann predictions according to Eq. (\protect\ref{Boltzmann}). Panels
(a), (b) correspond to the symmetric potential, $V_{0}=\pm u$, and (b), (d)
for asymmetric one, $V_{0}=u$.}
\label{fig:weak}
\end{figure*}

Let us now turn to the case of weak scattering. We consider two cases, the
symmetric potential, $V_{0}=\pm u,$ and the asymmetric one, $V_{0}=u.$
Figures \ref{fig:DOS} (b), (d) show the DOS and the electron density $n=n(E)$
for the case of $n_{i}=2\%$. As in the case of the strong scatterers, the
calculated dependencies are practically identical for random and correlated
distributions of impurities. However, in contrast to the case of the strong
scatterers, the DOS does not form the impurity band close to the Dirac point.

Figure \ref{fig:weak} shows the time-dependent diffusion coefficients $D(t)$
and the density dependence of the conductivity for $n_{i}=3\%$ with random
and correlated distributions of disorders. In contrast to the case of the
strong scattering potential, in the present case the diffusion coefficient
reaches its maximum at the same time $t\approx 120$ fs independent of the
impurity concentration. The density dependence of the conductivity, $\sigma
=\sigma (n),$ for symmetric scatterers is shown in Fig. \ref{fig:weak} (b).
In agreement with the Boltzmann predictions, Eq. (\ref{Boltzmann}), the
calculated Kubo conductivity is independent of the carrier density, even
though its calculated value differs by a factor of $\sim 3$ from the
corresponding theoretical predictions. The calculated $\sigma $ is
independent on $n$ for all densities except those in the vicinity of the
Dirac point where the conductivity shows a pronounced dip. This can be
explained by the fact that close to the Dirac point the graphene sheets is
in the electron-hole puddle density regime with strong potential variations
comparable to the Fermi energy of electrons. Because Eq. (\ref{Boltzmann})
is not valid in such the regime, the calculated density strongly deviates
from the predictions of Eq. (\ref{Boltzmann}). Figure \ref{fig:weak} (b)
also shows the the conductivity for the case of correlated impurities with $%
r_{0}=0.6r_{0\max }$ and $r_{0}=r_{0\max }.$ Similarly to the case of strong
scatterers the correlation in the impurity distribution practically does not
affect the conductivity of the system at hand.

The density dependence the conductivity, $\sigma =\sigma (n),$ for the case
of asymmetric potential, $V_{0}=u,$ is presented in Fig. \ref{fig:weak} (d).
It has two features that are different from the case of the symmetric
potential. First, for negative electron densities, the calculated
conductivity is independent of $n$, which is consistent with the Boltzmann
predictions, Eq. (\ref{Boltzmann}). However, for positive densities, the
conductivity shows a sublinear density dependence distinct from Eq. (\ref%
{Boltzmann}). Second, for the case of correlated impurities the conductivity
is increased by up to 30\% in the region of negative densities (when $\sigma
\approx $const$)$, whereas $\sigma $ is not affected by the correlations for
positive densities where the behavior is sublinear. It should be stressed
that the Boltzmann theory predicts the same density dependence regardless
whether the potential $V_{0}$ symmetric or not, whereas our numerical
calculations show a clear difference between these two cases. Note that
calculations with the asymmetric potential, $V_{0}=-u$, show the density
dependence of the conductivity which is mirror-symmetric to the case of $%
V_{0}=u.$

\begin{figure*}[tbh]
\includegraphics[keepaspectratio, width=1.8\columnwidth]{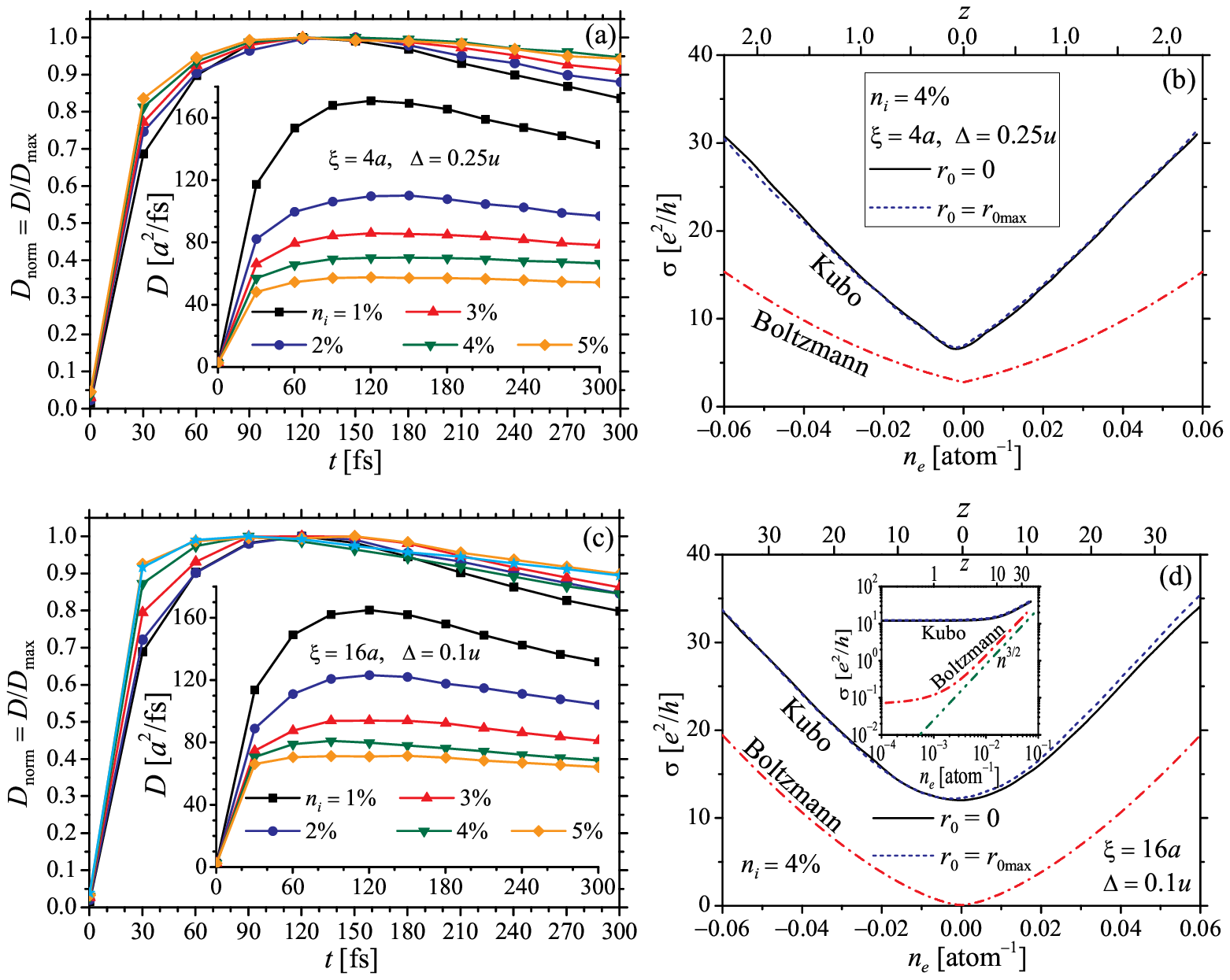}
\caption{(Color online) (a), (c) The diffusion coefficient $D=D(t)$ for the
Gaussian impurities with different concentrations $n_{i}$; $E=0.2u$. (b),
(d) The conductivity $\protect\sigma $ as functions of the relative electron
density $n_{e}$ (the number of electrons per C-atom) for random and
correlated Gaussian impurities, $n_{i}=4\%$. A dot-dashed line shows the
Boltzmann predictions according to Eq. (\protect\ref{BoltzmannGauss}). Inset
in (d) is plotted in the logarithmic scale. Panels (a), (b) and (c), (d)
correspond to $\protect\xi =4a$ and $\protect\xi =16a$ respectively.}
\label{fig:Gauss}
\end{figure*}

\subsection{\protect\bigskip Long-range Gaussian impurities}

Let us now turn to the case of the long-range Gaussian potential, Eq. (\ref%
{Gauss}). The graphene conductivity calculated within the the standard
Boltzmann approach in the Born approximation reads\cite{Adam,Klos,Vasko},

\begin{equation}
\sigma =\frac{4e^{2}}{h}\frac{\pi n\xi ^{2}e^{\pi n\xi ^{2}}}{KI_{1}(\pi
n\xi ^{2})}\propto \left\{
\begin{array}{c}
\text{const,}\;|z|\ll 1\text{,} \\
n^{3/2},\;|z|\gg 1.\text{ }%
\end{array}%
\right.  \label{BoltzmannGauss}
\end{equation}%
where $z\equiv \pi n\xi ^{2}=\left( \frac{2\pi \xi }{\lambda }\right) $, $%
I_{1}$ is the modified Bessel function and $K\approx 40.5n_{i}(\Delta
/u)^{2}(\xi /\sqrt{3}a)^{4}$ with $a$\textbf{\ }being the carbon-carbon
distance and $\lambda $ being the Fermi wavelength. The condition $|z|\ll 1$
describe the case of quantum scattering when the Fermi wavelength is larger
than the effective screening length, $\lambda \gg \xi ,$ while the opposite
condition $|z|\gg 1$ corresponds to the case of classical scattering, $%
\lambda \ll \xi .$ Because the semiclassical approach predicts two distinct
regimes of conductivity, in our numerical calculations we explore the
parameter space corresponding to these two regimes. We consider $\xi =4a$
and $\xi =16a,$ which in the considered density interval $|n_{e}|\lesssim
0.06$ correspond respectively to $|z|\lesssim 2,$ and $|z|\lesssim 35.$ Note
that the regime $|z|\lesssim 1$ is appropriate to realistic experimental
samples, whereas the carrier density $|z|\gtrsim 1$ is too high to be
achieved in the experiment \cite{Klos,Xu2011}.

Figures \ref{fig:DOS} (c), (d) show the DOS and the electron density $n=n(E)$
for the case of $n_{i}=2\%$. The Gaussian potential significantly smooths
the density of states especially in the region of the van-Hove
singularities. As in the case of the short-range scatterers, the calculated
dependencies are practically identical for random and correlated
distributions of impurities.

Figure \ref{fig:Gauss} shows the time-dependent diffusion coefficients $D(t)$
and the conductivity $\sigma =\sigma (n)$ for $n_{i}=4\%$ with random and
correlated distributions of disorders. As in the case of the weak
short-range disorder in the present case the diffusion coefficient reaches
its maximum value $D_{\max }$ at the same time for all impurity
concentrations studied $1\%\leq n_{i}\leq 5\%$ ($t=130$ fs respectively $110$
fs for $\xi =4a$ respectively $16a$).

The density dependence of the conductivity for $\xi =4a$ is shown in Fig. %
\ref{fig:Gauss} (b) in the density interval $|z|\lesssim 2.$ The
conductivity exhibits the linear density dependence which well extends into
the classical regime $|z|>1$. This is in agreement with previous numerical
calculations performed using the Green's function technique\cite%
{Lewenkopf,Klos}, also showing linear (or quasi-linear) density dependence
of the conductivity. On the contrary, the standard Boltzmann approach
predicts that for $|z|\ll 1$ the conductivity is independent on the electron
density, $\sigma =$const. For the present case of the Gaussian potential it
has been argued that the standard Boltzmann approach (leading to $\sigma =$%
const) is not valid because the Born approximation is not well justified.%
\cite{Klos}

Figure \ref{fig:Gauss} (d) shows the conductivity for the case of Gaussian
disorder for $\xi =16a$ in the density interval $|z|\lesssim 35.$ Even in
this case the calculated conductivity shows the linear density dependence, $%
\sigma \propto n.$ For $|z|\gg 1$ the Boltzmann approach predicts the
superlinear dependence $\sigma \propto n^{3/2}$ (this is clearly discernible
in the logarithmic scale as shown in the inset of Fig. \ref{fig:Gauss} (d)).
We however do not reach such high densities to reproduce this asymptote.
(This regime of the high densities was analyzed in Ref. \cite{Adam}).

Figures \ref{fig:Gauss} (b) and (d) also shows the conductivity for the case
of the correlated impurity distribution for $\xi =4a$ and $\xi =16a.$ Even
for the maximal correlation between impurities, $r_{0}=r_{0\max }$ the
correlation practically does not affect the conductivity. This represents
the central result of the present section.

\subsection{Comparison to the experiment}

Let us now compare our numerical data with available experimental results
and theoretical predictions. Recently, Yan and Fuhrer reported conductivity
measurements of potassium doped graphene.\cite{YanFuhrer}. They found that
with the increase of the temperature the conductivity of doped graphene
increases by up to a factor of 2. They attributed this effect to the
enhancement of the spatial correlation between potassium ions due to the
mutual Coulomb repulsion which, according to the recent theory of Li \textit{%
et al.}\cite{Li} leads to the conductivity enhancement. Our exact numerical
calculations however do not support this conclusion. Indeed, we demonstrated
that the spatial correlations of charged impurities (modeled by the
long-range Gaussian potential) has no effect on the conductivity of graphene
sheets. Our results therefore suggest that the enhancement of the
conductivity with increase of the temperature in experiments of Yan and
Fuhrer \textit{}\cite{YanFuhrer} is most likely caused by other
factors not related to the correlations of impurities. Moreover, as no
direct evidence of the spatial correlation of the potassium ions was
presented in the above article, it is not clear whether impurities in
experimental samples are correlated at the first place, and whether this
correlation (if any) increases with the temperature.

Our exact numerical results do not apparently support theoretical
predictions of Li \textit{et al.}\cite{Li} that the correlations in the
impurity positions lead to the enhancement of the conductivity. As mentioned
before, it has been shown that for the long-range Gaussian potential in a
parameter range corresponding to realistic systems the standard Boltzmann
predictions are in quantitative and qualitative disagreement with the exact
numerical results.\cite{Klos} The reason for that is the utilization of the
Born approximation which relies on the unperturbed wave functions of ideal
graphene without impurities. This is justified only for the case of weak
scattering and apparently fails for the long-range Gaussian scatterers in
the parameter range corresponding to realistic systems. (For the discussion
of the Born approximation for graphene and bilayer graphene see Refs. \cite%
{Ferreira,Klos,Xu2011}). This explains the discrepancy the between the exact
results and predictions of Li \textit{et al.}\cite{Li} which are essentially
based on the standard Boltzmann approach in the Born approximation.

\section{Conclusions}

Using an efficient time-dependent real space Kubo formalism we performed
numerical studies of conductivity of large graphene sheets with random and
correlated distribution of disorder. In order to describe realistic disorder
we used models of the short-range scattering potential (appropriate for
adatoms covalently bound to graphene) and the long-range Gaussian potential
(appropriate for screened charged impurities on graphene and/or dielectric
surface). The calculations for the uncorrelated potentials are compared to
the corresponding predictions based on the semiclassical Boltzmann approach
and to exact numerical calculations performed by different methods.

We find that for the most important, experimentally relevant cases of
disorder, namely the strong short-range potential and the long-range
Gaussian potential, the correlation in the distribution of disorder does not
affect the conductivity of the graphene sheets as compared to the case when
disorder is distributed randomly. This represent the main result of our
study. We find that the correlations lead to the enhancement of the
conductivity only for the case of the weak short-range potential and only
when the potential is asymmetric, i.e. $V=V_{0}$ or $V=-V_{0}.$ No
enhancement of the conductivity is found for the symmetric weak short-range
potential, $V=\pm V_{0}.$

Using our results we analyze the recent experiment of Yan and Fuhrer \textit{}\cite{YanFuhrer}\cite{Li} where the temperature increase of the
conductivity was attributed to the enhancement in the spatial correlation of
the adsorbed potassium ions. Our numerical findings do not sustain this
interpretation and our results strongly suggest that the enhancement of the
conductivity reported in the above study is most likely caused by other
factors not related to the correlations of impurities.

Our numerical calculations do not support theoretical predictions of Li
\textit{et al.}\cite{Li} that the correlations in the impurity positions for
the long-range potential lead to the enhancement of the conductivity. We
attribute this to the utilization of the standard Boltzmann approach within
the Born approximation which is not justified for the case of a long-range
potential in the parameter range corresponding to realistic systems.

\begin{acknowledgments}
The authors greatly appreciate discussions with A.-P. Jauho and M. Brandbyge concerning the time-dependent Kubo formalism.
A financial support from the Swedish Institute is greatly acknowledged.
\end{acknowledgments}

\appendix*

\section{Numerical calculation of the dc conductivity $\protect\sigma $ on
the basis of the time-dependent real space Kubo method}

\subsection{Calculation of the DOS $\protect\rho (E)$ and the time-dependent
diffusion coefficient $D(E,t)$}

Numerical calculation of the dc conductivity $\sigma $ requires computation
of the DOS $\rho (E)=$Tr$\left[ \delta (E-\hat{H})\right] $ and the
time-dependent diffusion coefficient $D(E,t),$ Eq. (\ref{Dx}). Let us start
with an algorithm for calculation of $\rho (E).$

Express the density of states $\rho (E)$ as a sum over the local densities
of states $\rho (E)=\sum_{i}^{N}\rho _{i}(E),$ where the summation is
performed over all the sites of graphene lattice $N$. In its turn, the local
density of states is given by the imaginary part of the diagonal elements of
the Greens function, $\rho _{i}(E)=-\frac{1}{\pi }\Im G_{ii}(E+i\varepsilon
),$ where a small $\varepsilon \rightarrow 0$ is introduced in order to
smooth the peaks in the DOS.

There is an efficient numerical algorithm for calculation of the diagonal
elements $G_{ii}$. It first starts with the tridiagonalization procedure
when the Hamiltonian is reduced to the tridiagonalized form by passing to a
new basis. Then the first diagonal element of the Green's function, $G_{11},$
is computed using the standard continued fraction technique. The details of
this algorithm are presented in Appendix B. Note that this algorithm scales
as the number of sites $N,$ with the most time-consuming part being the
triagonalization procedure. Having calculated the local density of states
for the first site, $\rho _{1}(E),$ one can in principle repeat similar
calculations for all remaining $N-1$ sites in order to recover the total DOS
$\rho (E).$ This would apparently require total computation efforts that
scale as $N^{2}.$ For a disordered systems treated in the present paper one
can adopt a different way to evaluate $\rho (E)$ that still keeps the total
computation efforts of the order of $N.$ It relies on the fact that a
sufficiently large subsystem of the total system has the same density of
states as the original one. Our computation of $\rho (E)$ proceeds then as
follows.

Choose a subsystem occupying $M$ sites of the graphene lattice. Construct a
random state $|\psi _{r}\rangle $ extended over these $M$ sites,
\begin{equation}
|\psi _{\text{ran}}\rangle =\frac{1}{\sqrt{M}}\sum_{i}e^{2i\pi \alpha
_{i}}|i\rangle ,  \label{psiRandom}
\end{equation}%
with $\alpha _{i}$ being the random phase in the interval $[0,1],$ $%
|i\rangle =c_{i}^{\dagger }|0\rangle ,$ and the summation is extended over
the sites of the chosen subsystem. An example of such a random state is
shown in Fig. \ref{fig:diff} (b). Transform the original tight-binding
Hamiltonian Eq. (\ref{H}) defined in the basis $\{r_{i}\}$ by passing on to
a new basis as follows. Choose the first element of the new basis as $%
|1\}=|\psi _{\text{ran}}\rangle $, where $|\psi _{\text{ran}}\rangle $ is
the random state defined according to Eq. (\ref{psiRandom}). Tridiagonalize
the Hamiltonian as prescribed in Appendix and calculate $\rho _{1}(E)=-\frac{%
1}{\pi }\Im G_{11}(E+i\varepsilon )$ using the continued fraction technique.
Repeat this procedure, if needed, for different distributions of disorder
and average $\rho _{1}(E)$ over these disorder realizations. The calculated
value of $\rho _{1}(E)$ corresponds to the DOS per carbon atom of the system
at hand. Note that calculation of the remaining $N-1$ matrix elements $G_{ii}
$ is not needed such that the total computational efforts still scale as $N.$

Calculation of the trace Tr$\left[ \left( \hat{X}_{H}(t)-\hat{X}(0)\right)
^{2}\delta (E-\hat{H})\right] $ in the expression for $D(E,t)$ in Eq. (\ref%
{Dx}) is performed in a similar way. It can be shown\cite{Markussen} that
this trace can be represented as a sum of the local densities of states Tr$%
\left[ \left( \hat{X}_{H}(t)-\hat{X}(0)\right) ^{2}\delta (E-\hat{H})\right]
=\sum_{i}^{N}\rho _{i}(E,t)$ for the functions $\Psi _{i}(t)$, where
\begin{equation}
\rho _{i}(E,t)=\sum_{i=1}^{N}\left\langle \Psi _{i}(t)|\delta (E-\hat{H}%
)|\Psi _{i}(t)\right\rangle ,  \label{Tr}
\end{equation}%
and
\begin{equation}
|\Psi _{i}(t)\rangle =\hat{X}\hat{U}|\psi _{i}\rangle -\hat{U}\left( \hat{X}%
|\psi _{i}\rangle \right) ,  \label{Psii(t)}
\end{equation}%
where $\hat{X}$ is the position operator (i.e. the $x$-coordinate), $\hat{U}$
is the time evolution operator and $\{\psi _{i}\}$ is the orthogonal basis
set (corresponding to e.g. the eigenstates of the Hamiltonian Eq. (\ref{H}%
)). Next, we pass on to the new basis setting its first element $|1\}=|\Psi
_{i}(t)\rangle ,$ where $|\Psi _{i}(t)\rangle $ is given by Eq. (\ref%
{Psii(t)}) where $|\psi _{i}\rangle =|\psi _{\text{ran}}\rangle $ is chosen
as the random state defined by Eq. (\ref{psiRandom}). Time evolution of the
random initial state in Eq. (\ref{Psii(t)}) is calculated by means of the
Chebyshev method as described below. We then reduce the Hamiltonian to the
tridiagonalized form and use the continued fraction technique to calculate
local density of states $\rho _{1}(E,t)$ in the new bases. As argued above,
for a sufficiently large random state, this local density of state well
approximates the density of states of the whole system.

\subsection{Exact solution of the time-dependent Schr\"{o}dinger equation by
means of the Chebyshev method}

In this appendix we present an efficient method for solution of the
time-dependent Schr\"{o}dinger equation based on the expansion of the time
evolution operator in an orthogonal set of Chebyshev polynomials.

We start from the time-dependent Schr\"{o}dinger equation complemented by
the initial condition
\begin{equation}
\hat{H}\left\vert \psi (t)\right\rangle =i\hbar \frac{\partial }{\partial t}%
\left\langle \psi (t)\right\vert ,\text{ }\left\vert \psi (t=0)\right\rangle
=\left\vert \psi _{0}\right\rangle ,  \label{timeSchrod}
\end{equation}%
with $\hat{H}$ being the tight-binding time-independent Hamiltonian, Eq. (%
\ref{H}). Its formal solution can be expressed via the time-evolution
operator $\hat{U}(t)$,%
\begin{equation}
\left\vert \psi (t)\right\rangle =\hat{U}(t)\left\vert \psi
_{0}\right\rangle ,\text{ }\hat{U}(t)=e^{-\frac{i}{\hbar }\hat{H}(t)}.
\label{psi(t)}
\end{equation}%
In order to expand $\hat{U}(t)$ in a set of the Chebyshev polynomials $%
T_{n}(x)$ (which are defined in the interval $x\in \lbrack -1;1])$, we we
first renormalize the Hamiltonian such that its spectrum lies in the above
interval,
\begin{equation}
\hat{H}_{\text{norm}}=\frac{2\hat{H}-(E_{\max }+E_{\min })\hat{I}}{E_{\max
}-E_{\min }},  \label{Hnorm}
\end{equation}%
where $E_{\max }$ and $E_{\min }$ are the largest and the smallest
eigenvalues of the original Hamiltonian Eq. (\ref{H}). (In order to
calculate $E_{\max }$ and $E_{\min }$ we use a computational routine that
estimates the largest/smallest eigenvalues of the operator without
calculation of all the eigenvalues).

Expanding $\hat{U}(t)$ in Chebyshev polynomials in Eq. (\ref{psi(t)}) we
obtain for the wave function,
\begin{equation}
\left\vert \psi (t)\right\rangle =\sum\limits_{n=0}^{\infty
}c_{n}(t)\left\vert \Phi _{n}\right\rangle ,  \label{expansion}
\end{equation}%
where the functions $\left\vert \Phi _{n}\right\rangle =T_{n}(\hat{H}_{\text{%
norm}})\left\vert \psi _{0}\right\rangle $ are calculated using the
recurrence relations for the Chebyshev polynomials,
\begin{equation}
\left\vert \Phi _{n+1}\right\rangle =2\hat{H}_{\text{norm}}\left\vert \Phi
_{n}\right\rangle -\left\vert \Phi _{n-1}\right\rangle ,  \label{recurrence}
\end{equation}%
with $\left\vert \Phi _{0}\right\rangle =T_{0}(\hat{H}_{\text{norm}%
})\left\vert \psi _{0}\right\rangle =\left\vert \psi _{0}\right\rangle $.
The expansion coefficients $c_{n}(t)$ are calculated making use of the
orthogonality relation for the Chebyshev polynomials%
\begin{equation}
c_{n}(t)=2e^{-i\frac{(E_{\max }+E_{\min })t}{2\hbar }}(-i)^{n}J_{n}\left(
\frac{E_{\max }-E_{\min }}{2\hbar }t\right) .  \label{c(t)}
\end{equation}%
For large $t$ the expansion coefficients $c_{n}(t)$ become exponentially
small. This leads to the fast convergence of the expansion series Eq. (\ref%
{expansion}), and makes the Chebyshev method very efficient for calculation
of the temporal dynamics. For instance, in our simulations we take 10000
iterations for 300 fs of elapsed time.

\subsection{Continued fraction technique and tridiagonalization of the
Hamiltonian matrix}

\subsubsection{Continued fraction technique}

Consider a Hamiltonian matrix in a tridiagonal form,
\begin{equation}
\hat{H}_{\text{tri}}=\left(
\begin{array}{ccccccc}
\alpha _{1} & \beta _{1} & 0 & \cdots & \cdots & \cdots & \cdots \\
\beta _{1} & \alpha _{2} & \beta _{2} & 0 & \cdots & \cdots & \cdots \\
0 & \beta _{2} & \alpha _{3} & \beta _{3} & 0 & \cdots & \cdots \\
\vdots & 0 & \beta _{3} & \ddots & \ddots & \ddots & \cdots \\
\vdots & \vdots & 0 & \ddots & \ddots & \ddots & \ddots \\
\vdots & \vdots & \vdots & \ddots & \ddots & \ddots & \beta _{N-1} \\
\vdots & \vdots & \vdots & \vdots & \ddots & \beta _{N-1} & \alpha _{N}%
\end{array}%
\right)  \label{Htri}
\end{equation}%
The continued fraction technique provides an efficient way to calculate the
first diagonal element $G_{11}$ of the Greens function $\hat{G}=(E\hat{I}-%
\hat{H}_{\text{tri}})^{-1}$ without a need for computing the whole Greens
function and all the eigenvalues/eigenfunctions of the Hamiltonian.

Let us denote $\lambda _{i}=G_{1i}$. From the definition of the Greens
function we obtain,
\begin{equation}
(E-\alpha _{i})\lambda _{i}-\beta _{i-1}\lambda _{i-1}-\beta _{i}\lambda
_{i+1}=0,\;2\leq i\leq N-1  \label{lambda}
\end{equation}%
with $(E-\alpha _{N})\lambda _{N}-\beta _{N-1}\lambda _{N-1}=0$ and $%
(E-\alpha _{1})\lambda _{1}-\beta _{1}\lambda _{2}=1$. Expressing
sequentially $\lambda _{N}$ by $\lambda _{N-1}$, $\lambda _{N-1}$ via $%
\lambda _{N-2}$ and $\lambda _{N-3}$ we express $G_{11}$ as a continued
fraction:
\begin{equation}
G_{11}=\frac{1}{E-\alpha _{1}-\frac{\beta _{1}^{2}}{E-\alpha _{2}-\frac{%
\beta _{2}^{2}}{\frac{\ddots }{E-\alpha _{M}-\beta _{M}^{2}\Sigma (E)}}}},
\label{G11}
\end{equation}%
In the above equation we truncated the order-$N$ continued fraction at the
fraction $M<N$ by introducing the self-energy $\Sigma (E),$%
\begin{equation}
\Sigma (E)=\frac{1}{E-\alpha _{M}-\frac{\beta _{M}^{2}}{E-\alpha _{M}-\frac{%
\beta _{M}^{2}}{\ddots }}}=\frac{1}{E-\alpha _{M}-\beta _{M}^{2}\Sigma (E)},
\label{selfEnergy1}
\end{equation}%
that includes all the remaining terms $M+1\leq i\leq N$. Solving Eq. (\ref%
{selfEnergy1}) one easily obtains
\begin{equation}
\Sigma (E)=\frac{E-\alpha _{M}-i\sqrt{4\beta _{M}^{2}-(E-\alpha _{M})^{2}}}{%
2\beta _{M}^{2}}.  \label{selfEnergy2}
\end{equation}%
The number of terms $M$ included in the summation in Eq. (\ref{G11}) is
determined from the condition for the convergence of $G_{11}.$ For instance,
in our calculations with the $N\times N$ Hamiltonian matrixes with $%
N=1.7\times 10^{6}$ and $6.8\times 10^{6}$ it is sufficient to choose $%
M\approx 2000$ and $4000$ respectively.

\subsubsection{Tridiagonalization of the Hamiltonian matrix}

In order to utilize the continued fraction technique to calculate $G_{11}$
the Hamiltonian should be transformed to the tridiagonalized form, Eq. (\ref%
{Htri}). This is done by constructing a new orthogonal basis as described
below. We start by selecting the first basis vector $|1\}$. We use curled
brackets $|i\}$ to denote the new basis vectors and straight brackets $%
|i\rangle $ to denote the old ones. If the tridiagolization is performed in
order to find the local density if states on the $i$-th site of the system
at hand, the first basis vector is selected as $|1\}=$ $\left\vert
i\right\rangle $, where $|i\rangle =c_{i}^{\dagger }|0\rangle .$ In most our
calculations we select the first basis vector as the random state occupying $%
M$ sites, $|1\}=|\psi _{\text{ran}}\rangle ,$ with $|\psi _{\text{ran}%
}\rangle $ being given by Eq. (\ref{psiRandom}).

We require that the Hamiltonian in the new basis be of the tridiagonal form (%
\ref{Htri}). By operating $\hat{H}|i\}$ we arrive to the following
equations,
\begin{subequations}
\label{basis}
\begin{eqnarray}
\hat{H}|1\} &=&\alpha _{1}|1\}+\beta _{1}|2\},  \label{Basis_a} \\
\hat{H}|i\} &=&\beta _{i-1}|i-1\}+\alpha _{i}|i\}+\beta _{i}|i+1\},\text{ }%
2\leq i\leq N-2,  \label{Basis_B} \\
\hat{H}|N\} &=&\beta _{N-1}|N-1\}+\alpha _{N}|N\}.  \label{Basis_c}
\end{eqnarray}%
Using Eq. (\ref{Basis_a}) and the orthogonality relation $\{1|2\}=0$ we
obtain the second basis vector and the matrix elements $\alpha _{1}$ and $%
\beta _{1}$,
\end{subequations}
\begin{eqnarray}
|2\} &=&\frac{1}{\sqrt{C_{2}}}\left( \hat{H}|1\}-\alpha _{1}|1\}\right) ,
\label{Basis_2} \\
\alpha _{1} &=&\{1|\hat{H}|1\},\;\beta _{1}=\{2|\hat{H}|1\},  \notag
\end{eqnarray}%
where the normalization coefficient $C_{2}$ (as well as all other
normalization coefficients $C_{i},2\leq i\leq N)$ are obtained from the
normalization requirement $\{i|i\}=1.$

We then proceed to Eq. (\ref{Basis_B}) and recursively calculate the basis
vectors $|i\},2\leq i\leq N-2$, and corresponding matrix elements $\alpha
_{i}$ and $\beta _{i},$
\begin{eqnarray}
|i+1\} &=&\frac{1}{\sqrt{C_{i+1}}}\left( \hat{H}|i\}-\beta
_{i-1}|i-1\}-\alpha _{i}|i\}\right) ,  \label{Basis_i} \\
\alpha _{i} &=&\{i|\hat{H}|i\},\;\beta _{i+1}=\{i+1|\hat{H}|1\},\text{ }%
2\leq i\leq N-2  \notag
\end{eqnarray}%
Finally, from Eq. (\ref{Basis_c}) we obtain
\begin{equation*}
|N\}=\frac{1}{\sqrt{C_{N}}}\left( \hat{H}|N\}-\alpha _{N}|N\}\right) ,\text{
}\alpha _{N}=\{N|\hat{H}|N\},
\end{equation*}%
which concludes the tridiagonalization procedure.

\begin{figure*}[tbh]
\includegraphics[keepaspectratio, width=2.0\columnwidth]{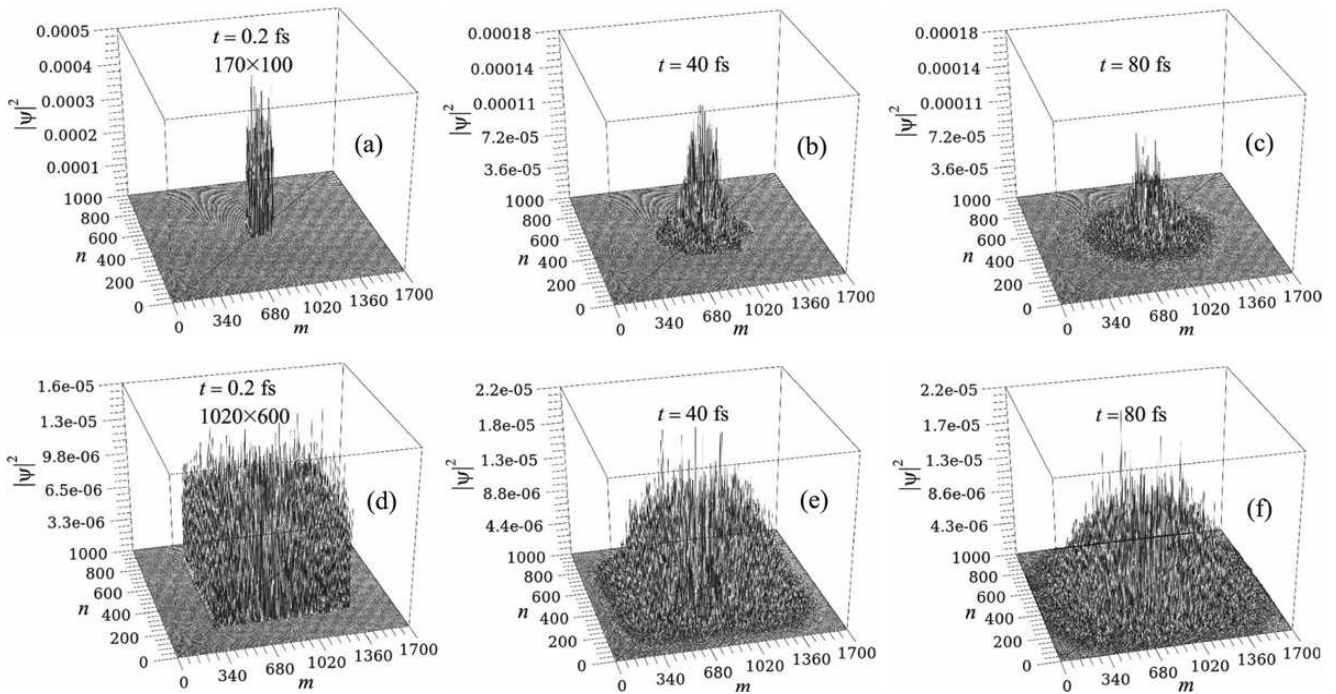}
\caption{Propagation of wave packets of two different initial sizes: 170$%
\times $100 and 1020$\times $600 on a graphene sheet with random
distribution of the short range strong impurities $V=37u$, $n_{i}=1\%$.}
\label{fig:wavePackets2}
\end{figure*}

\subsection{Role of the initial wave packet size in the averaging over
impurity configurations}

\begin{figure*}[tbh]
\includegraphics[keepaspectratio, width=2.0\columnwidth]{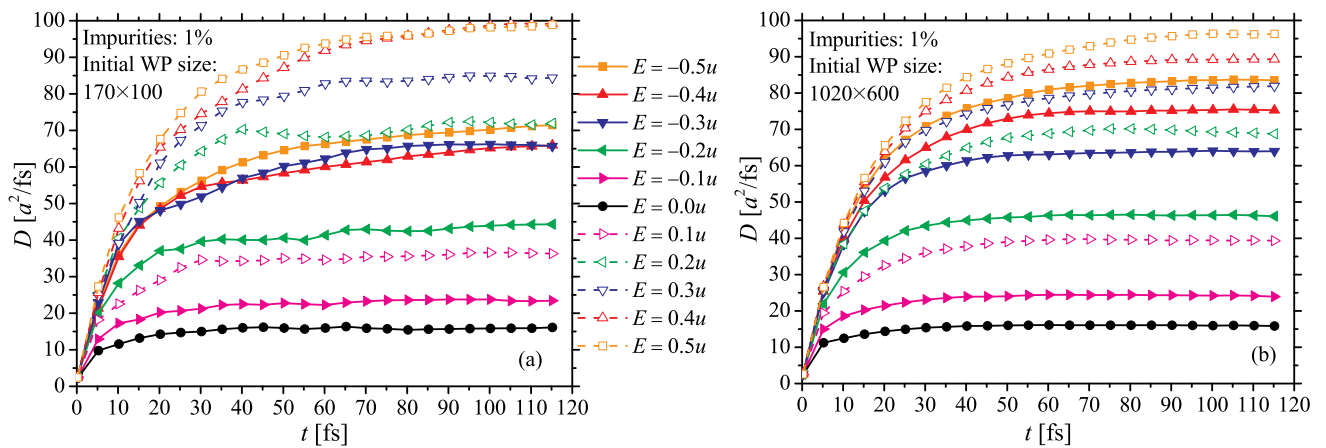}
\caption{(Color online) Time-dependent diffusion coefficient $D(t)$ at
different energies $E$ for the wave packets (WP) of Fig. \protect\ref%
{fig:wavePackets2}. }
\label{fig:DE}
\end{figure*}

\begin{figure*}[tbh]
\includegraphics[keepaspectratio, width=2.05\columnwidth]{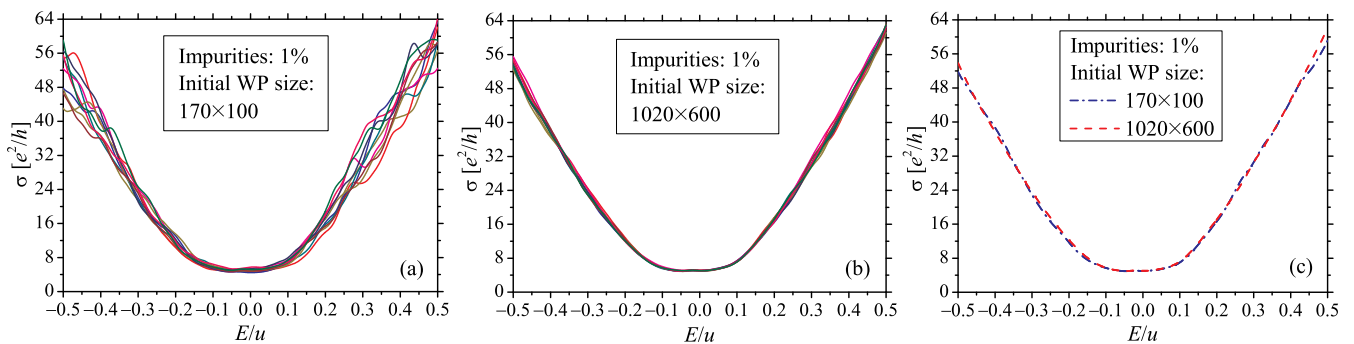}
\caption{(Color online) (a), (b) Conductivity vs. energy for the wave packet
(WP) of Fig. \protect\ref{fig:wavePackets2} for ten different impurity
configurations. (c) Averaged values of conductivities for the wave packets
from (a) and (b).}
\label{fig:sigmaComparison}
\end{figure*}
For a given configuration of impurities, the conductivity of the system at
hand can be sensitive to details of the potential and thus can vary from one
impurity realization to another. The conductivity therefore should be
averaged by e.g. performing many calculations for different impurity
distributions. Within the present time-dependent Kubo approach the averaging
can be done in a much more efficient way by simply increasing a size of the
initial random state without need for many different calculations for
different impurity realizations. In this section we investigate how one can
achieve an efficient averaging of the conductivity by varying the size of
the wave packet.

Figure \ref{fig:wavePackets2} shows a temporal evolution of two random
initial states $|\psi _{\text{ran}}\rangle $ (Eq. (\ref{psiRandom})) of
different sizes, 170$\times $100 and 1020$\times $600 respectively. The
corresponding time-dependent diffusion coefficients $D(t)$ are shown in Fig. %
\ref{fig:wavePackets2} for different electron energies $E$. For the case of
the smaller packet the calculated temporal dependencies show fluctuations
caused by the interference within the wave packet. In contrast, for the case
of the larger packet these fluctuations are efficiently averaged out and $%
D(t)$ exhibit a smooth monotonic behavior for all considered energies. These
self-averaging features of the wave packets temporal dynamics manifest
themselves in the density dependencies of the conductivity $\sigma =\sigma
(n).$ Figure \ref{fig:sigmaComparison} (a),(b) show the dependencies $\sigma
=\sigma (n)$ for the above wave packets for different impurity realizations
(Note that the concentration of impurities $n_{i}$ in all cases is the
same.) For the case of the smaller wave packet the conductivity exhibits
significant fluctuations, whereas for the case of the larger wave packet the
conductivity curves practically coincide. It is important to stress that the
averaged values of the conductivity are the same in both cases, see Fig. \ref%
{fig:sigmaComparison} (c). Thus, in most our calculations we choose the wave
packet to be sufficiently large (typically 1020$\times $600) such that no
additional averaging over impurity realization is needed.

It is noteworthy that a larger wave packet apparently reaches the boundary
of the computational domain earlier than the smaller one, c.f. Fig. \ref%
{fig:wavePackets2} (c) and (f). The size of the computational domain should
be therefore sufficiently large, such that the maximum value of the
diffusion coefficient, Eq. (\ref{a}), is reached before the wave packet hits
the boundaries. In our calculations we used the honeycomb lattices of the
sizes 1700$\times $1000 and 3400$\times $2000 sites (corresponding to 210$%
\times $210 and 420$\times $420 nm$^{2}$ respectively).


\begin{thebibliography}{99}
\bibitem{T-matrix} A graphene lattice with an adatom situated at the $j$-th
site can be described by the Hamiltonian $\hat{H}=-u\sum_{i,i^{\prime
}}c_{i}^{\dagger }c_{i^{\prime }}-u_{ad}\left( c_{j}^{\dagger
}a_{j}+c_{j}^{\dagger }a_{j}\right) +\epsilon _{ad}c_{j}^{\dagger }c_{j},$%
where the operators $c_{j}$ and $a_{j}$ correspond to respectively graphene
lattice and the adatom, $u$ is defined in Eq. (\ref{H}), $u_{ad}$ is the
hopping integral between the adatom and the carbon atom of the graphene
lattice at the site $j,$ and $\epsilon _{ad}$ is the on-site energy at the
adatom. Using the T-matrix theory one can show that this Hamiltonian can be
reduced to Eq. (\ref{H}) where the strength of the on-site potential in Eq. (%
\ref{delta}) is\cite{Robinson,Wehling,Ihnatsenka,Ferreira} $%
V_{0}=u_{ad}^{2}/(E-\epsilon _{ad}).$ With typical values of $u_{ad}$ and $%
\epsilon _{ad}$ provided by the ab-initio calculations the effective on-site
potentials in the vicinity of the Dirac point in the range $|V_{0}|/u\approx
10-80$.\cite{Ferreira}
\bibitem{derivative} Note that the general definition of the diffusion
coefficient is $D_{x}(E,t)=\frac{d\left\langle \Delta \hat{X}%
^{2}(E,t)\right\rangle }{dt}.$ This definition becomes equivalent to Eq. (%
\ref{a}) in a classical diffusive regime when $D$ is independent of time and
the mean spreading of a wave packet is liniarly proportional to time, $%
\left\langle \Delta \hat{X}^{2}(E,t)\right\rangle \propto t.$ However, even
in the case when $D$ depends on $t$ both definitions lead to qualitatively
same results.\cite{Lherbier12} This justify utilization of Eq. (\ref{a})
which is less demanting computationally.
\bibitem{ElDensity} In our numerical calculations we use relative electron
densities in the range $|n_{e}|\lesssim 0.06$ ($|n|\lesssim 2.3\times
10^{14} $cm$^{-2}$). This is larger than those typically used in
experiments, $|n_{e}^{\exp }|\lesssim 5\times 10^{-5}$ ( $|n^{\exp
}|\lesssim 2\times 10^{11}$cm$^{-2}$). We use larger electron densitiy
interval because we model conductivity using impurity densities $%
n_{i}\lesssim 5\%$ ($n_{i}\lesssim 2\times 10^{11}$cm$^{-2}$) which are
larger than those in typical experimental samples, $n_{i}^{\exp }\lesssim
0.05\%$ ($n_{i}^{\exp }\approx 2\times 10^{11}$ cm$^{-2}$).\cite{PeresReview}
This is due to the fact that with the realistic impurity densities one would
need to perform calculations on graphene sheets with thouthands of millions
of atom in order to achieve the diffusive transport regime, which exceeds
the present computation capabilities.
\end{thebibliography}
\end{document}